\documentclass[]{spie}  

\usepackage{amsmath,amsfonts,amssymb}
\usepackage{graphicx}
\usepackage[colorlinks=true, allcolors=blue]{hyperref}

\title{Detection of isolated specular reflection for calibration of cloud thermodynamic phase estimation with quantum parametric mode sorting LIDAR}

\author[a]{Richard J. Murchie}
\author[a]{Dolf Huybrechts}
\author[a]{Aaron Strangfeld}
\author[b]{Mateusz P. Mrozowski}
\affil[a]{ESTEC, European Space Agency, Keplerlaan 1, 2201 AZ, Noordwijk, The Netherlands.}
\affil[b]{Fraunhofer CAP, 99 George Street, Glasgow, G1 1RD, United Kingdom.}

\authorinfo{Corresponding author: richard.murchie@esa.int}

\begin{document} 
\maketitle

\begin{abstract}
Specular reflection can be problematic for the determination of the cloud thermodynamic phase for near-nadir-pointing space LIDARs. A LIDAR system biased towards the specular contribution for backscatter, if near-concurrent to a conventional LIDAR, could calibrate the measurements required for cloud phase determination. One such system which shows promise for this is quantum parametric mode sorting (QPMS) LIDAR. Through a non-linear interaction and time-frequency mode selectivity, this system demonstrates in-band noise-rejection beyond what linear noise filtering can provide. This level of noise-rejection means the signal strength can be minimised, therefore biasing the specular contribution to the return signal. Here we provide a theoretical model of QPMS LIDAR applied to this scenario to instruct its feasibility.
\end{abstract}


\section{INTRODUCTION}
\label{sec:intro}  
A space LIDAR system with its transmitter and receiver operating at near-nadir can experience anomalously enhanced backscatter due to the specular reflection of hydrometeors. In particular, horizontally-orientated ice crystals (HOICs) are a large contributor to the enhanced backscatter. Atmospheric microphysics models require accurate knowledge of the cloud thermodynamic phase. A basic approach for estimation of cloud phase is via thresholded sectors according to a function of the layer-integrated depolarisation $\delta$ and backscatter $\gamma$ \cite{Hu09}. However, the specular backscatter from HOICs in an ice cloud can confuse this estimation algorithm. As for example, their presence can result in an anomalously high backscatter. Furthermore, single-bounce reflection off HOICs and spherical water droplets retain polarisation, whereas reflection off randomly-orientated ice crystals (ROICs) can cause depolarisation. Therefore, the combination of both the high backscatter and low depolarisation mimics the characteristics of a liquid water cloud according to the thresholded cloud phase sectors, rather than an ice cloud laden with ROICs \cite{Hu07}. This can lead to an erroneous classification of cloud phase and an under-reporting of the incidence of ice clouds \cite{Zhu25}.

Retrieval algorithms have been developed to reduce misclassification from specular contamination. The CALIPSO Version 4 cloud phase algorithm combines the aforementioned threshold sectoring with a spatial coherence test to discriminate HOIC backscatter against water cloud particles or ROIC backscatter. This spatial coherence test involves measuring an adjacent footprint of the same cloud system. Whichever footprint has the higher backscatter will show a lower (negatively-correlated) depolarisation if it is an ice cloud dominated with HOICs. Otherwise, if the higher backscatter footprint shows a higher (positively-correlated) depolarisation then it indicates either an ice cloud with ROICs or a water cloud. Cloud phase estimation is then further supplemented with data from centroid temperature, colour ratio, and view angle gates, operating on top of improved 532~nm and 1064~nm calibrations \cite{Hu09,Avery20,Liu19}. However, residual biases remain in complicated scenes such as mixed-phase and multilayer clouds. Thus, while this algorithmic compensation is demonstrably effective, it remains a statistical correction that cannot always resolve the underlying ambiguity, motivating interest in alternative measurement concepts.

A separate LIDAR system which is biased to only register specular contributions for its return signal could assist a conventional LIDAR system by instructing the influence of specular backscatter. In turn, the specular contribution of the conventional LIDAR return signal could be accounted for, thereby reducing erroneous classification of cloud phase. Specular reflection is mainly a single-bounce scattering event which concentrates most of its reflected irradiance onto a small solid angle, unlike diffuse (Lambertian-like) scattering or multiple-scattering (volume) events. Therefore, we strive for a LIDAR system which can operate at a minimised signal strength: such that the return signal is biased towards only specular reflections. Moreover, we strive for a system which preferentially passes singly-scattered contributions to the return signal. One such system that can provide both of these requirements is quantum parametric sorting (QPMS) LIDAR \cite{Shahverdi17}. This separate LIDAR system will transmit its signal in-between the transmission of the conventional system: thereby ensuring that the existing receiver and transmitter optics of the platform can easily be reused.

QPMS LIDAR operates via a non-linear three-wave mixing process in a $\chi^{(2)}$ medium, typically implemented in a waveguide. This non-linear process is an example of quantum frequency conversion (QFC) and in particular, as we are performing up-conversion, it is a sum-frequency generation (SFG) process. The system is operated near the edge of phase-matching, which enables spatio-temporal mode-selective frequency up-conversion. In this scheme, a strong pump beam and a signal beam, both at 1064~nm, are incident on the non-linear medium. The signal beam interrogates the environment, whereas the up-converted output (at 532~nm) of the non-linear interaction is conventionally referred to as the idler, which is subsequently detected. The key advantage of QPMS LIDAR is that the up-conversion process is inherently mode-selective, allowing it to reject background noise with an efficiency that surpasses the theoretical limits of the ideal time-frequency filter \cite{Shahverdi17}. Furthermore, because multiple-scattering events can degrade the spatio-temporal mode structure of the signal, the system preferentially up-converts singly scattered photons. This feature allows QPMS LIDAR to maintain high fidelity in environments where multiply scattered light would otherwise dominate, thus enabling low-noise and high-precision ranging under challenging conditions.

   \begin{figure} [ht]
   \begin{center}
   \begin{tabular}{c} 
   \includegraphics{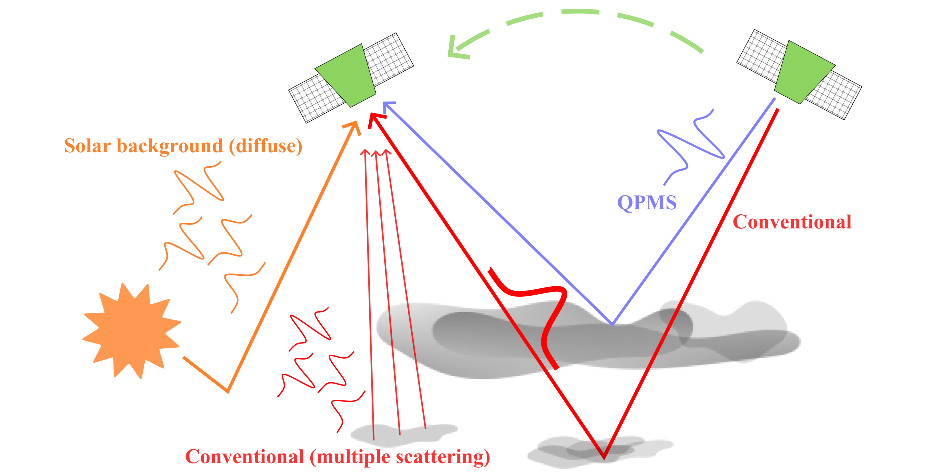}
   \end{tabular}
   \end{center}
   \caption[]{A schematic of the QPMS LIDAR system alongside the noise sources we consider.
    \label{fig:system} }
   \end{figure} 

For this paper, we consider that the initial spatial-temporal mode structure of our signal beam is a fundamental Gaussian spatial mode and a Hermite-Gaussian (HG) temporal mode (TM). Moreover, we will consider the QFC regime where our signal and pump beams have the same group velocity: known as the single sideband velocity matched (SSVM) regime \cite{ReddyD13}. As will be seen, double propagation through the atmosphere, reflection, and coupling into our receiver optics will degrade the spatio-temporal mode structure of our signal beam: which negatively affects the performance of QPMS LIDAR. Figure~\ref{fig:system} is a schematic of the QPMS LIDAR operating in conjunction with a conventional LIDAR. The noise sources are visualised by the lines not corresponding to the QPMS LIDAR backscatter incident on the receiver. Furthermore, the TM structuring of the system is apparent: the QPMS LIDAR has its TM as the third order HG mode and the specular return of the conventional LIDAR is the fundamental Gaussian (zeroth order HG mode). The multiply scattered conventional LIDAR backscatter and the diffusive solar background noise are a combination of HG modes for their TM structure. Lastly, it is illustrated that the transmission of the conventional and QPMS LIDARs pulses are out of sync.

In Sect.~\ref{sect:TM_prelim} we give the preliminary details and formalism for the spatio-temporal mode structure of our light. Section~\ref{sect:prop} we provide a model for the double atmospheric propagation and specular reflection of our signal beam, from which we show how the spatio-temporal mode structure evolves within this model. For Sect.~\ref{sect:noise} we model the expected noise sources in this system. Section~\ref{sect:LIDAR} consolidates the propagation, noise, signal, and QFC process analysis into a more holistic model of the QPMS LIDAR system. We then derive the detection statistics for a range of operating conditions. This section also discusses how QPMS LIDAR could provide calibration for a near-concurrent conventional LIDAR system. Lastly, in Sect.~\ref{sect:conclusion} we discuss the results and conclusions of this conceptual study.

\section{Spatio-temporal modes preliminary}
\label{sect:TM_prelim}
We consider that the spatial aspect of our field amplitudes (for signal and pump) are initially described by the fundamental transverse Gaussian mode \cite{Andrews05}. We denote this transverse field amplitude in Cartesian coordinates $\mathbf{R}=(x,y,z)$, (where $z$ is the longitudinal axis) in the plane of the transmitter ($z=0$) as \begin{equation}
    \psi(x,y,0)=\sqrt{\frac{2}{\pi W_0^2}}e^{-\frac{1}{W_0^2}(x^2+y^2)}, 
\end{equation} where $W_0$ is the effective beam radius such that the amplitude falls to $e^{-1}$ of its value on the centre of beam axis ($x,y=0$) and $k=\frac{2\pi}{\lambda}$ is the optical wave-number. We do not need to explicitly consider the spatial mode in terms of the quantum theory of light. This is because we will only use the results of the spatial loss calculation due to our finite receiver aperture size and reflection. In other words, other than an attenuation pre-factor, the spatial mode evolution will not be directly involved in our calculation for the QFC process.

To describe the TM structure of our beams we will move onto the quantum mechanical description of light \cite{Brecht15}. The canonical quantisation of the electromagnetic field for a monochromatic mode with angular frequency $\omega$ yields its associated bosonic creation operator $\hat{a}^\dagger(\omega)$, which adheres to the commutation relation $[\hat{a}(\omega),\hat{a}^\dagger(\omega^{'})]=\delta(\omega,\omega^{'})$, where $\hat{a}(\omega)=(\hat{a}^\dagger(\omega))^\dagger$ is the hermitian conjugate of the creation operator and $\delta(\omega,\omega^{'})$ is the Dirac delta function. The creation operator can be interpreted as the creation of a photon of energy in the corresponding mode. We assume a fixed polarisation and that our transverse field spatial mode is in a beam-like geometry. A complex (and normalised) spectral amplitude $f_j(\omega)$ defines a TM indexed by $j$. We can also describe the TM in the time-domain, with $f_j(t)$ as the Fourier transform of $f_j(\omega)$. Thus, the broadband creation operator which creates a single-photon in this TM, in the frequency-domain and time-domain respectively is \begin{equation}\hat{A}^\dagger_j=\frac{1}{2\pi}\int d\omega f_j(\omega)\hat{a}^\dagger(\omega)=\int d t f_j(t) \hat{a}^\dagger(t),\end{equation} where $\hat{a}(t)$ follows the usual bosonic commutation relations. The broadband creation operator also obeys a bosonic commutation relation $[\hat{A}_i,\hat{A}^\dagger_j]=\delta_{ij}$, where $\delta_{ij}$ is the Kronecker delta. What is interesting about TMs is that while they fully overlap spatially, time-frequency and in polarisation, they are orthogonal over a frequency integral (time integral) \begin{equation}
    \frac{1}{2\pi}\int d\omega f_i^*(\omega)f_j(\omega)=\int dt f_i^*(t)f_j(t)=\delta_{ij}.
\end{equation} This orthogonality means that they are suitable for use as a basis. Therefore, we consider a family of orthogonal complex spectral amplitudes: the HG functions. For a TM described by an HG of order $j$ it is defined as, \begin{equation}
    f_j(\omega;\omega_0)=\frac{\mathcal{N}}{\sqrt{n!\sqrt{\pi}2^n \sigma}}H_j\left(\frac{\Delta\omega}{\sigma}\right)e^{-\frac{(\Delta\omega)^2}{2\sigma^2}},
\end{equation} where $\mathcal{N}=\sqrt{2\pi}$ is the normalisation factor, $H_j(x)$ is a Hermite polynomial of order $j$, $\Delta\omega=\omega-\omega_0$ with $\omega_0$ as the central frequency of the mode and $\sigma$ is the spectral width of the amplitude set by $e^{-1}$. 
To avoid non-linear effects during atmospheric propagation we set the pulse (temporal) width as $T_p=200$~ps (based off the zeroth order HG TM) \cite{Kasparian08}, this value instructs the spectral width $\sigma$, whereby pulse width roughly scales as $\sqrt{j}$ with the HG mode order $j>0$. The details on how the pump and signal can be manipulated into an HG TM of our choice is given in Ref.~\cite{Kowligy14}. For brevity, we will denote a quantum state with a TM described by an HG function of order $j$ as $\mathrm{TM}(j)$. Moreover, we shall model our signal beam to be initially in the fundamental Gaussian spatial mode with the Glauber coherent state, excited by a TM($j$), which has the quantum state \cite{Raymer20} \begin{equation}
    \vert \alpha\rangle_{\mathrm{TM(j)}}=\mathrm{exp}(\alpha \hat{A}^\dagger_j-\alpha^*\hat{A}_j)\vert\mathrm{vac}\rangle,
\end{equation} where $\alpha$ is the complex amplitude of the coherent state and $\vert \mathrm{vac}\rangle$ is the frequency continuum vacuum state. The mean photon number of this state is readily calculated $\bar{n}=\vert \alpha\vert^2$.

\section{Atmospheric propagation and reflection}
\label{sect:prop}
We assume that the solution to the scalar wave equation is separable by its spectral-temporal and spatial components, thus its spatial and temporal aspects do not interact during propagation and reflection and so we can model these dynamics separably.
\subsection{Spatial}
This section quantifies the spatial loss due to divergence/diffraction and the finite aperture size. Due to the high-altitude region of interest we shall neglect the effect turbulence has on our spatial mode: further details for modelling the effect of atmospheric turbulence for a spatial mode is detailed in Ref.~\cite{Wang20}

A monochromatic beam spatial amplitude evolves in free-space according to the scalar paraxial equation \begin{equation}
\nabla^2_\mathrm{T}\psi(\mathbf{R})+i2k\partial_z\psi(\mathbf{R})=0,\label{eq:stochastic_waveeqn}
\end{equation} where $\nabla^2_\mathrm{T}$ is the transverse Laplacian operator. We can directly solve the scalar paraxial equation with the propagated Gaussian spatial amplitude at the far-field position in the optical axis $z$ in polar coordinates ($r^2=x^2+y^2$) as\begin{equation}
    \psi(r,z)=-i\sqrt{\frac{2}{\pi W_0^2}}\frac{W_0}{W(z)}e^{\frac{-r^2}{W(z)^2}}e^{ikz},
\end{equation} where $W(z)=W_0\sqrt{1+(\frac{z}{z_R})^2}$ is the beam-waist at distance $z$ and $z_R=\frac{\pi W_0^2}{\lambda}$ is the Rayleigh range \cite{Andrews05}. Following on from this we model reflection off a target by a multiplication of a reflectivity mask (aperture function) onto our spatial amplitude. The specifics of this mask and how it relates to the proportion of HOICs in the region of interest are detailed in Sect.~\ref{sect:HOIC_geometry}. The final leg (from reflector to receiver) is in the far-field and we can use the theory of Fraunhofer diffraction to calculate the spatial amplitude of our field upon the receiver plane.

From our analysis of the effect propagation and reflection has upon our spatial mode we calculate the spatial mode mismatch coefficient \cite{Wang20}. This coefficient encapsulates the attenuation of our signal beam due to spatial loss. We define the spatial mode mismatch coefficient as \begin{equation} C_{\mathrm{SMM}}=\vert \int\int \psi_0^*(x,y,2L) A(x,y)\psi_\mathrm{1}(x,y,2L) dx \;dy\vert^2,\label{eq:C_SMM}
\end{equation} where $\psi_0$ is the spatial field unaffected by reflection and receiver aperture clipping, $\psi_\mathrm{1}$ is the spatial mode reflected off our target and clipped by the receiver aperture function $A(x,y)$. 
\subsubsection{Reflector characteristics}
\label{sect:HOIC_geometry}
Accurate modelling of the backscatter from ice crystals in cirrus is a non-trivial problem, in part due to the impracticality of solving Maxwell's equations and the limitations of geometric optics for large size scatterers, whereby physical optics approximations are employed \cite{Kono17}. Instead, we use a simple model by considering our specular reflector as a circular planar mirror (with a reflectivity factor $r_\mathrm{spec}=0.05$). This model will give a notion of the operating conditions with respect to the proportion of specular reflector covering the beam footprint. However, it will not model the changes to polarisation, HOIC tilt distribution and diffraction effects due to backscatter from an ensemble of separate HOICs, for example. Our field incident on the reflection plane is multiplied by a mask $r(x,y)=\sqrt{r_\mathrm{spec}}R(x,y)$, where $R(x,y)$ expresses the geometry of the reflector. This mask $r(x,y)$ encodes the effect a reflector has on our field incident on the reflection plane. In polar coordinates ($\tilde{r}=\sqrt{x^2+y^2}$) we express the circular aperture with radius $R$ as \begin{equation}R(x,y)=R(\tilde{r})=\begin{cases}
    1:\tilde{r}\leq R,\\ 0:\tilde{r}> R.
\end{cases}\end{equation}
\subsection{Temporal}
The following analysis instructs the resulting TM structure of our signal beam when it is incident at the SFG waveguide. For modelling the TM evolution we can neglect atmospheric turbulence as there is negligible frequency dependence of turbulent variations of the refractive index \cite{Banakh14}. We follow the procedure for TM evolution through the atmosphere as in Ref.~\cite{Wang22} where they treat the atmosphere as a lossless dispersive linear optical medium. Atmospheric loss due to absorption/scattering is included later as a separate attenuation factor. The approach from Ref.~\cite{Wang22} segments the atmosphere into $1$km thick layers to treat the altitude-dependent refractive index, where the atmosphere tops out at $100$~km. As we are probing for HOICs, we limit the lowest layer of the atmosphere in our dispersion calculations to an altitude of $4$km. Moreover, specular reflection does not affect our TM structure: there is experimental evidence that TMs are robust against small-angle scattering \cite{Zhu21}.

The complex spectral amplitude, for TM($j$) after suffering a double trip of atmospheric-induced dispersion is \begin{equation}
    f^{2 L}_j(\omega,\omega_0)=f_j(\omega,\omega_0)\;\mathrm{exp}\left(-i2\sum^{100}_{q=5} k_q(\omega)L_q\right),\label{eq:dispersion}
\end{equation} where $k_q(\omega)=\langle n(\lambda,h_{q-1})\rangle\frac{\omega}{c}$ is the propagation factor for the $q^{\mathrm{th}}$ layer of the atmosphere, $\langle n(\lambda,h_{q-1})\rangle$ is the mean refractive index for the wavelength $\lambda$ and altitude $h_{q-1}=(q-1)\times10^3$ (m) and $c$ is the speed of light in vacuum. Removal of the group delay relative to the original pulse and the group delay dispersion will help counter the dispersion experienced. A Taylor expansion (around the central frequency $\omega_0$ allows us to quantify the effect of removal of the group delay (first order dispersion) and group delay dispersion (second order dispersion). Regardless, higher-order dispersion will still affect the TM structure as it is not compensated for. We refer to the dispersion-compensated complex spectral amplitude as \begin{equation}
    \tilde{f}^{2 L}_j(\omega,\omega_0)=f^{2 L}_j(\omega,\omega_0)\;\mathrm{exp}\left(i(\omega-\omega_0)\sum^{100}_{q=5}\frac{d k_q(\omega)}{d\omega}\bigg\vert_{\omega_0} 2 L_q+i(\omega-\omega_0)^2\sum^{100}_{q=5}\frac{d^2 k_q(\omega)}{d\omega^2}\bigg\vert_{\omega_0}L_q\right).
\end{equation}Following on from this we can calculate the overlap coefficient of the complex spectral amplitude TM($k$) to the dispersed/compensated complex spectral amplitude (denoted by $j$). It shows how retained a particular TM is during atmospheric propagation: this will feed into how much signal loss is anticipated. The overlap coefficient is $\vert c^{\mathrm{TM}}_{k,j}\vert^2$, where \begin{equation}
    c^{\mathrm{TM}}_{k,j}=\frac{1}{2\pi}\int d\omega f^*_k(\omega,\omega_0) \tilde{f}^{2 L}_j(\omega,\omega_0).\label{eq:overlap}
\end{equation}

\begin{figure} [ht]
   \begin{center}
   \begin{tabular}{c} 
   \includegraphics[height=5cm]{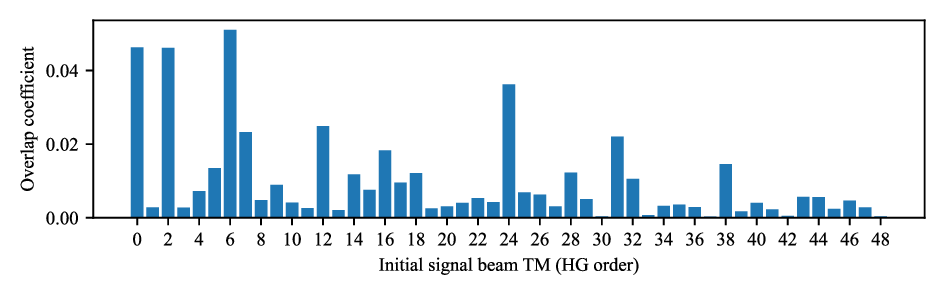}
   \end{tabular}
   \end{center}
   \caption[]{The TM overlap coefficient as a function of the TM HG order, after double propagation through the atmosphere and reflection for the dispersed complex spectral amplitude.
    \label{fig:TM_structure} }
   \end{figure}

Figure~\ref{fig:TM_structure} shows the TM overlap coefficient $\vert c^\mathrm{TM}_{k,j}\vert^2$ as a function of the HG order of the initial signal beam TM, for the dispersed complex spectral amplitude. It is clear that each complex spectral amplitude is scattered away from being concentrated on their initial HG TM. However, for a narrowband pulse of temporal width $T_p=200$~ps, any atmospheric dispersion is successfully compensated for by accounting for group delay and group delay dispersion. For shorter pulses, the higher-order effects of atmospheric dispersion cause the TM structure to severely degrade even with the partial dispersion compensation.

\section{Background noise} 
\label{sect:noise}
\subsection{Diffusive and multiple-scattering QPMS contribution}
\label{sect:diffusive_noise}
The QPMS LIDAR signal strength is minimised to ensure the diffusive scattering influence is negligible. Moreover, multiple-scattering often induces dephasing which can damage the TM structure and thus will hamper up-conversion: this means the contribution of multiple-scattering is limited in the QPMS return signal \cite{Zhu21}.

One form of diffusive scattering is via the mechanism of Rayleigh scattering. Any loss from Rayleigh scattering is incorporated by the atmospheric transmissivity factor $T_{\mathrm{atm}}$ and we can neglect any increase to our QPMS return signal, due to the minimised signal strength we focus on. The other form of diffusive scattering we consider is Lambertian scattering. An order of magnitude estimate suffices to instruct the signal strength which would yield negligible Lambertian backscatter. A beam experiences free-space divergence from transmitter to the reflector, which can be quantified by the half-angle divergence $\theta_\mathrm{div}=\frac{\lambda}{\pi W_0}$. In the small-angle approximation, this results in a beam footprint radius of $r_\mathrm{fp}=\mathrm{tan}(\theta_\mathrm{div})z$. From this there is the fraction that the reflector covers this beam footprint $f$. If we assume that the reflector is circular then there is the half-angle divergence induced by the reflector $\theta_\mathrm{diffr}\approx0.618\frac{\lambda}{R}$ where $R$ is the radius of the reflector. Hence, we approximate the effective half-angle divergence as $\theta_\mathrm{eff}\approx\sqrt{\theta_\mathrm{div}^2+\theta_\mathrm{diffr}^2}$. In the far-field, assuming the specular and Lambertian reflectors are the same size and cover the same fraction of the beam footprint we have the specular to Lambertian ratio as $\frac{1}{\theta_\mathrm{eff}^2}$. For a reflector size of $50~\mu $m, a nominal initial beam waist $W_0=6.77$~mm (from the $50~\mu$rad half-angle beam divergence of the CALIOP LIDAR \cite{Winker17}) and central wavelength $\lambda=1064~$nm the specular to Lambertian ratio is $5782$. As the reflector size becomes smaller, the reflection-induced diffraction dominates over the free-space divergence and the specular to Lambertian ratio reduces. If the reflector size becomes comparable to the wavelength we can no longer consider it as specular or Lambertian scattering and must consider it via physical optics techniques or solving Maxwell's equations. In such scenarios there will be much more spatial loss, due to the lack of a narrow specular lobe directed towards the receiver. Hence, for a minimised signal system such as our QPMS LIDAR it will not be able to detect such small-scale reflectors. 

\subsection{Solar background noise}
\label{sect:solar_noise}
Solar background noise is incoherent and stationary (over the relevant time-scales of the non-linear interaction) and it can be modelled as multi-mode thermal light. Given a set polarisation we can write the electric field amplitude of a thermal field as linear combination of TMs weighted by a Gaussian random variable $A_i$, for each TM($i$)  \cite{Su19}. The expectation value $\langle A^*_iA_i\rangle$ corresponds to the mean photon number $\bar{n}_{\mathrm{B,i}}$ of TM$(i)$ \cite{Banaszek19}. Following this, we use the model by Ref.~\cite{Degnan02} for the total solar background noise mean photon number over the bandwidth and temporal duration set, excluding the surface backscatter aspect. We define the total solar background noise mean photon number $\bar{n}_\mathrm{B}=\frac{\lambda N_\lambda}{hc}A_{\mathrm{det}} \Delta \lambda \;T_p$, where $A_\mathrm{det}$ is the receiver area, $h$ is Planck's constant,  $N_\lambda=7.49\times10^{-10}$ is the solar flux ($\mathrm{W/m^2/nm}$) for a central wavelength $\lambda=1064~$nm and a ground albedo $0.3$, for a receiver field-of-view $\Omega=1.33\times10^{-8}$~sr and a solar radiance value from MODTRAN \cite{Berk14}. Moreover, $\Delta\lambda=0.00708$~nm is the spectral bandwidth in wavelength, calculated from the Gaussian transform-limited pulse width $T_p$ of our TM(0) pulse. Hence, the total solar background mean photon number is approximately $\bar{n}_\mathrm{B}=0.0045$,

We relate the total solar background mean photon number to the solar background mean photon number for TM ($i$) as $\bar{n}_\mathrm{B}=\sum^{N-1}_{i=0}\bar{n}_\mathrm{B,i}$. There is no preference of TM for the (white-light) solar background noise and under the approximation of truncating the span of TMs to $N=40$, we have that for each TM($i$) its mean photon number is $\bar{n}_\mathrm{B,i}=\frac{\bar{n}_\mathrm{B}}{N}$. Hence, the background noise mean photon number for each TM($i$) is $\bar{n}_\mathrm{B,i}=1.125\times10^{-4}$.
\subsection{Conventional LIDAR}
We base our model of a conventional LIDAR system off the CALIOP LIDAR system, but we shall make a couple of adjustments \cite{Winker17}. For example, instead of its specified pulse width $20$~ns we set it match the QPMS LIDAR system at $200$~ps, this ensures the CALIOP pulse temporal amplitude is already in the HG TM basis. We set the HG TM of the conventional LIDAR pulse to be TM(0): the fundamental Gaussian. Moreover, we set the spatial mode to be the fundamental transverse Gaussian. As discussed earlier, we model the state of light corresponding to LIDAR to be the Glauber coherent state and we set the conventional LIDAR amplitude as $\alpha_c=7.66\times10^7$. Furthermore, the conventional system will experience multiple-scattering returns, on account of its signal strength being orders of magnitude stronger than QPMS. We define the amplitude of the multiple-scattering return as $\delta\alpha_c$.

\section{LIDAR system}
\label{sect:LIDAR}

\subsection{Detection statistics}
We seek to solve for what detected QPMS signal strength (mean photon number $\bar{n}_\mathrm{S}$) is required to confidently measure isolated specular reflection. Therefore, we consider and compare two states incident upon our SFG waveguide for analysing the detection statistics: a signal-only state and a noise-only state (which includes contributions from the solar background and conventional LIDAR). The derivation of these quantum states is given in Appendix~\ref{sect:quantum_state}. The signal-only state $\hat{\rho}_\mathrm{S}$ is given by Eq.~\ref{eq:signal_only}, where we set $\mathrm{TM}(3)$. The noise-only state $\hat{\rho}_\mathrm{B}$ is given by Eq.~\ref{eq:noise_only}, where we set TM$(0)$. The signal-only state has the attenuation factor $\eta$ and the noise-only state has the attenuation factor $\eta_c$. The differing attenuation factors is due to the off-sync operation of the QPMS and conventional system: hence different reflectors causing specular reflection to concur during the QPMS detection window. The attenuation factor $\eta/\eta_c$ is constituted by \begin{equation}
    \eta=\;C_\mathrm{SMM}\;T^2_\mathrm{atm},
\end{equation}where $T_\mathrm{atm}=0.856$ is the atmospheric transmissivity for a wavelength of $\lambda=1064~$nm \cite{Berk14}. The attenuation factor $\eta$ is dependent on the geometry of the reflector. Even though both states experience the same dispersion, it results in different scattering on each HG TM, thus the $c^\mathrm{TM}_{k,j}$ coefficients differ. We include the multiple-scattering of the conventional LIDAR return with amplitude $\delta\alpha_c$ for each HG TM in the noise state Eq.~\ref{eq:noise_only}. This results in the transformation of the noise-only coherent state amplitude as $\sqrt{\eta_c}\;c^{\mathrm{TM}}_{k,(j=0)}\alpha_c+i\sqrt{1-\eta_c}\;\beta_k \to \sqrt{\eta_c}\;(c^{\mathrm{TM}}_{k,(j=0)}\alpha_c+\frac{\delta\alpha_c}{\sqrt{\eta_c}})+i\sqrt{1-\eta_c}\;\beta_k$. 

From the derivation of the SFG process in Appendix~\ref{sect:SFG} we express the photon-number statistics of the SFG output, whether it is the signal state or noise incident. Both states also experience the same SFG waveguide thus the decomposition coefficients $\lambda_n$ are the same. The mean photon number of the idler TM($n$) is \begin{equation}
    \bar{n}_{\mathrm{i:n}}=\langle \hat{b}_n^\dagger\hat{b}_n\rangle,
\end{equation} whereby the SFG output idler channel mean photon number is\begin{equation}
  \bar{n}_\mathrm{i}=\sum_n\bar{n}_{\mathrm{i:n}}. \label{eq:total_SFG_output_mean_photon}
\end{equation} Furthermore, the SFG output idler channel photon number variance is \begin{equation}
    \Delta\bar{n}_i=\sqrt{\sum_n \langle \hat{b}_n^\dagger\hat{b}_n\hat{b}^\dagger_n\hat{b}_n\rangle-\bar{n}_{i:n}^2}.\label{eq:total_SFG_output_variance}
\end{equation}
From the expectation values detailed in Appendix~\ref{sect:photon_number} we can calculate the strictly standardised mean difference (SSMD) for a range of conditions. This quantity will instruct the operating conditions where QPMS LIDAR can confidently measure specular reflection. As we want to avoid diffusive scattering influence and polluting the conventional LIDAR signal we focus on the minimum QPMS mean photon number $\bar{n}_\mathrm{S}$ which obtains a SSMD value of 2. The SSMD is defined as \begin{equation}
    \mathrm{SSMD}=\frac{\bar{n}_\mathrm{S}-\bar{n}_\mathrm{B}}{\sqrt{\Delta\bar{n}_\mathrm{S}^2+\Delta\bar{n}_\mathrm{B}^2}},
\end{equation} where $\bar{n}_\mathrm{S/B}$ is defined by Eq.~\ref{eq:total_SFG_output_mean_photon} corresponding to the quantum state $\hat{\rho}_\mathrm{S/B}$. Furthermore, the SFG output photon number variance $\Delta\bar{n}_\mathrm{S/B}$ defined by Eq.~\ref{eq:total_SFG_output_variance} corresponds to the quantum state $\hat{\rho}_\mathrm{S}/\hat{\rho}_\mathrm{B}$. Formally, for a set of system parameters (such as background noise and attenuation), we solve for $\mathrm{min}(\bar{n}_\mathrm{S})\vert \mathrm{SSMD}=2$. 

\begin{figure} [ht]
   \begin{center}
   \begin{tabular}{c} 
   \includegraphics{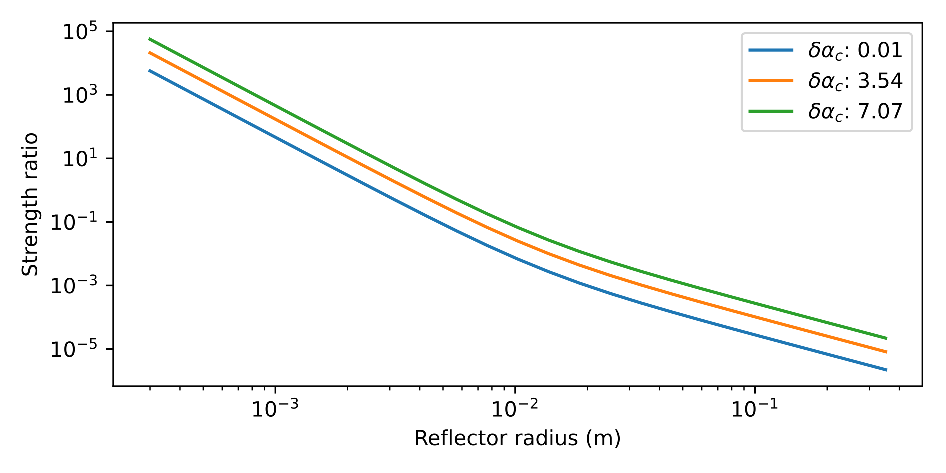}
   \end{tabular}
   \end{center}
   \caption[]{Strength ratio $\frac{\vert\alpha\vert^2}{\vert\alpha_c\vert^2}$ on the y-axis with reflector radius $R$ on the x-axis, for different amplitudes of multiple-scattering contribution $\delta\alpha_c$.
    \label{fig:detection} }
   \end{figure}

Figure~\ref{fig:detection} demonstrates the ratio of minimum QPMS LIDAR signal strength $\vert\alpha\vert^2$ to the conventional LIDAR signal strength $\vert\alpha_c\vert^2$ as a function of reflector radius $R$, for different amplitudes of multiple-scattering contribution $\delta\alpha_c$. Here, the minimum value of QPMS LIDAR signal strength is determined by the criterion of $\mathrm{min}(\bar{n}_\mathrm{S})\vert \mathrm{SSMD}=2$. From Fig.~\ref{fig:detection} it is clear that if the reflector radius is too small then the QPMS LIDAR signal strength will exceed that of the conventional system: this defeats the purpose of the system to be auxiliary to the conventional LIDAR. Furthermore, the conventional LIDAR multiply-scattered backscatter ($\propto\delta_c$) negatively affects the QPMS LIDAR performance more than the conventional LIDAR specular backscatter ($\propto\vert\alpha_c\vert^2)$. This is due to the multiply-scattered backscatter spanning many HG TMs rather than remaining in TM$(0)$. In Fig.~\ref{fig:detection} we set the specular reflector of the conventional LIDAR to have a radius of $100~\mu$m.

\subsection{Calibration with conventional LIDAR}
Since 1064 nm is a wavelength commonly used in satellite LIDAR, our QPMS LIDAR system, operating at the same wavelength, can take advantage of the established radiometric calibration methods developed for conventional LIDAR. Following this, from our model it is clear that the proportion of specular reflectors within the beam footprint in the region of interest is a microphysical parameter that can be extracted. This parameter is correlated with the proportion of HOICs. The proportion of HOICs can then be fed into a cloud phase algorithm decision tree such as in Ref.~\cite{Avery20}, particularly when cloud phase confidence is low.

Throughout our analysis we have neglected polarisation for brevity. However, a more advanced treatment would consider polarisation in the propagation and backscattering process. This would facilitate an estimation of the depolarisation ratio from the QPMS LIDAR measurements: important for the recognition of HOICs. The involvement of polarisation would thus necessitate two channels (for parallel and perpendicular polarisations) each with its own SFG waveguide and detector.

\section{Discussion}
This paper is a conceptual study of QPMS LIDAR operating in conjunction with a spaceborne conventional LIDAR. Its purpose is to preferentially detect specular backscatter. The specular contribution of backscatter from high-altitude probing correlates with the presence of HOICs. Therefore, a clean measurement of this specular contribution could help with the estimation of cloud thermodynamic phase: particularly in complicated scenes involving mixed-phase or multiple-layers. Our study suggests that under certain conditions (such as sufficient reflector size) QPMS LIDAR can confidently detect isolated specular backscatter from high-altitude clouds.

This study initiates with a discussion on the current techniques and limitations of cloud phase estimation. Whereby methods such as the spatial coherence test and supplementary data has been used to improve cloud phase classification. However, the presence of HOICs is often statistically inferred which inevitably introduces error. An alternative measurement concept, such as QPMS LIDAR, could help mitigate these errors. 

QPMS LIDAR operates via a non-linear three-wave mixing process known as sum-frequency generation (SFG). It involves spatial-temporal mode-selective up-conversion into the detection band. Due to this mode-selectivity: filtering has been demonstrated 11.3~dB beyond the ideal time-frequency filter and a 41~dB signal-to-noise ratio (SNR) advantage over direct detection has been reported \cite{Shahverdi17,Shahverdi18}. Such levels of filtering means we can minimise the QPMS LIDAR signal strength to avoid polluting the return signal with diffusive contributions. Moreover, multiple-scattering can degrade the TM structuring required for successful up-conversion; therefore, such multiply-scattered contributions are effectively filtered out. Consequently, the preference of QPMS LIDAR detecting singly-scattered backscatter also lends itself to the application of precision altimetry and ranging in turbid/strongly scattering media \cite{Maruca21,Zhu21}.

We provide a preliminary discussion about the spatial mode and TM of our signal field. Whereby we use the HG modes as the basis for our TMs. Following this, we develop a model for how the (double) atmospheric propagation and reflection affects the spatial and temporal aspects of our signal field, respectively. Our spatial model quantifies the signal attenuation expected due to propagation, the reflector characteristics and our finite receiver aperture size. The temporal model quantifies how the TM structuring of our field is degraded by dispersion in the atmosphere, we also quantify how dispersion-compensation can partially recover the initial TM structure. The noise sources (which includes multiply-scattered and specular reflection from the near-concurrent conventional LIDAR) are also treated within the spatial and temporal field evolution framework developed.

We combine the analysis thus far into a more holistic picture of a QPMS LIDAR system. From which we can calculate the detection statistics of the system (i.e. the photon number expectations post-SFG process). Following this, we solve for the minimum QPMS LIDAR signal strength which ensures confident detection for a given regime of noise and reflector size. However, too small of a reflector size and the QPMS LIDAR requires a signal strength greater than that of the conventional system. Moreover, our results show that the QPMS system performance is more vulnerable to excessive noise originating from the conventional LIDAR multiply-scattered backscatter rather than specular reflection.

Further work is required to demonstrate the feasibility of the application of QPMS LIDAR for the isolation of specular contributions to backscatter. For example, turbulence has been neglected in our model. The inclusion of turbulence will negatively affect the spatial coupling into the receiver aperture. The model for the reflector characteristics is simple and would benefit from a full physical optics treatment. Lastly, a treatment on how polarisation is affected by scattering is required: particularly due to its importance for cloud phase estimation.

\label{sect:conclusion}

\appendix

\section{Quantum state incident on the SFG waveguide}
\label{sect:quantum_state}
We want to calculate the quantum state of our attenuated and dispersion-compensated signal beam which is incident upon our receiver. This quantum state will also be mixed with background noise. Moreover, we wish to have the resultant state expressed in terms of TMs in the HG basis. This maps cleanly onto the Heisenberg picture SFG calculations later, as the TM annihilation operators correspond to our TM-decomposed state incident upon the receiver.  

The effect of dispersion and its partial compensation means only the higher-order dispersion terms on its Taylor expansion remain. This can be modelled as a unitary phase shift \begin{equation}
    \hat{U}=\text{exp}\left(-i\int d\omega \;\phi(\omega)\hat{a}^\dagger(\omega)\hat{a}(\omega)\right),
\end{equation} where $\phi(\omega)$ consists of the higher-order Taylor expansion terms in $2\sum^{100}_{q=5} k_q(\omega)L_q$ which are not cancelled by the dispersion compensation. In the Heisenberg picture this phase-shift evolves the displacement operator in a TM($j$) as \begin{align}
    \hat{U}\hat{D}(\alpha)_{\mathrm{TM(j)}}\hat{U}^\dagger&=\mathrm{exp}\left(\alpha\hat{A}_j^\dagger e^{i\phi(\omega)}+\alpha^*\hat{A}_j e^{-i\phi(\omega)} \right), \\
    &=\prod_k\mathrm{exp}\left(c^{\mathrm{TM}}_{k,j}(\alpha\hat{A}_k^\dagger+\alpha^*\hat{A}_k) \right), \\ &= \prod_k\hat{D}(c^{\mathrm{TM}}_{k,j}\alpha)_\mathrm{TM(k)}
\end{align} Here, we have performed a modal decomposition of $\hat{A}_je^{-i\phi(\omega)}$ into the HG basis. The result of the partial-dispersion is to scatter the coherent state in TM($j$) into a tensor product of TM-structured coherent states in the HG basis indexed by $k$, such that $\sum_k\vert c^{\mathrm{TM}}_{k,j}\vert^2=1$.

We parameterise the attenuation of our signal as $\eta$: this is a product of the spatial loss and the atmospheric (round-trip) attenuation. Attenuation of our signal beam and the mixing with noise can be modelled via a beamsplitter transformation acting independently onto each TM component of the signal and noise quantum states. The beamsplitter transformation, for TM($k$), can be expressed by the unitary operator \begin{equation}
    \hat{U}^\mathrm{BS}_\mathrm{TM(k)}=\mathrm{exp}\left(-i\;\mathrm{cos}^{-1}(\sqrt{\eta})(\hat{A}^\dagger_0\hat{A}_1+\hat{A}_0\hat{A}^\dagger_1)\right),
\end{equation} where $\hat{A}_0$ and $\hat{A}_1$ are annihilation operators for the same TM($k$), but for different input ports of the beamsplitter: by convention we use mode $0$ for our signal and mode $1$ for the noise/vacuum. Relabelling the output ports of the beamsplitter, from $2\to 0$ and $3\to 1$, we partial trace out mode $1$ as this represents the environment, which we do not have access to perform a measurement upon. The combined unitary to model attenuation and noise mixing is thus $\hat{U}^\mathrm{BS}=\prod_k \hat{U}^\mathrm{BS}_\mathrm{TM(k)}$.

We can express the output state $\hat{\rho}$ of the mixing of our signal light $\hat{\rho}_0$ and the background noise $\hat{\rho}_1$ as \begin{equation}
    \hat{\rho}=\mathrm{Tr}_1\left(\hat{U}^\mathrm{BS} \hat{\rho}_0\otimes\hat{\rho}_1\hat{U}^{\mathrm{BS}\dagger}\right).
\end{equation} If we have no background noise $\hat{\rho}_1=\bigotimes_k\vert 0\rangle\langle 0\vert_{\mathrm{TM(k)}}$ and thus our output state for this scenario is \begin{equation}
    \hat{\rho}=\bigotimes_k \vert \sqrt{\eta} c^{\mathrm{TM}}_{k,j}\alpha\rangle\langle \sqrt{\eta}c^{\mathrm{TM}}_{k,j}\alpha\vert_\mathrm{TM(k)}. \label{eq:signal_only}
\end{equation} When we have background noise in a thermal state for each TM($k$) $\hat{\rho}_1=\bigotimes_k \sum_n^\infty\frac{\bar{n}_k^n}{(\bar{n}_k+1)^{n+1}}\vert n\rangle\langle n\vert_\mathrm{TM(k)}$ then the output state is \begin{equation}
    \hat{\rho}=\bigotimes_k\frac{(1-\eta)}{\pi\bar{n}_k}\int d^2\beta_k \;\mathrm{exp}\left(\frac{-\vert \beta_k\vert^2(1-\eta)}{\bar{n}_k}\right)\vert \sqrt{\eta}\;c^{\mathrm{TM}}_{k,j}\alpha+i\sqrt{1-\eta}\;\beta_k\rangle\langle \sqrt{\eta}\;c^{\mathrm{TM}}_{k,j}\alpha+i\sqrt{1-\eta}\;\beta_k\vert_\mathrm{TM(k)}.\label{eq:noise_only}
\end{equation} We have scaled the TM($k$) thermal state mean photon number by $(1-\eta)$, so the signal beam attenuation does not affect the background noise statistics. Moreover, we have used the P-function to describe the thermal state in the coherent state basis, for analytic ease. In Sect.~\ref{sect:solar_noise} the thermal state mean photon number values per TM(k) is given.

\section{QFC process and detection}
QPMS LIDAR is underpinned by the QFC process which enables the TM selectivity and thus the non-linear filtering advantage. From the effective Hamiltonian of collinear SFG with a classical undepleted pump in Ref.~\cite{Eckstein11} the Heisenberg equations of motion can be calculated in the moving frame of the pump. These equations describe the propagation and interaction of single spatial mode light in a lossless SFG waveguide \begin{align}
    (\partial_z+\mu\partial_t)\hat{a}(z,t)&=i\kappa f(t)\hat{b}(z,t),\\
    (\partial_z+\nu\partial_t)\hat{b}(z,t)&=i\kappa^* f^*(t)\hat{a}(z,t),\label{eq:Heisenberg_eqns}
\end{align} where $\mu$ and $\nu$ are the inverse group velocities of the signal and idler waves, respectively, relative to the pump. We have $\hat{a}$ and $\hat{b}$ as the standard annihilation operators for the signal and idler modes, respectively. Moreover, $f(t)$ is the TM of the pump in the time-domain and $\kappa$ is a coefficient for the SFG interaction strength. When the idler mode is initially in the vacuum we can solve the Heisenberg equations of motion for the idler mode annihilation operator at the end of the SFG waveguide (of length $L_\mathrm{w}$) using the Green function formalism, thus \begin{equation}
    \hat{b}(L_\mathrm{w},t)=\int dt^{'}G_{\mathrm{is}}(t,t^{'})\hat{a}(0,t^{'})+\hat{\xi}(t),\label{eq:input_output}
\end{equation} where $\hat{\xi}(t)$ is the annihilation operator corresponding to the vacuum: as we neglect noise such as Raman noise \cite{ReddyD13,Huang13,Kuo13}. We set the input/output Schmidt modes of the Green function to be the HG TMs used thus far. In practice, the Schmidt modes of the Green function are determined by the conditions of the waveguide and the pump pulse shape, and will not in general be a one-to-one mapping from one HG TM to another HG TM. In other words, the output Schmidt mode of the idler may constitute a combination of signal HG TMs. Further details on this more general case are described in Appendix~\ref{sect:SFG_decomp}. The Green function for signal to idler is \begin{equation}
    G_{\mathrm{is}}(t,t^{'})=\sum^\infty_{n=0}\lambda_n \psi_n(t)\phi^*_n(t^{'}),
\end{equation} where $\phi_n(t)$ is the input signal HG TM, $\psi_n(t)$ is the output idler HG TM and $\lambda_n$ is the decomposition (Schmidt) coefficient. We can then decompose the signal to idler conversion relation Eq.~\ref{eq:input_output} into a signal to idler conversion relation by HG TM\begin{equation}
    \hat{b}_n=\lambda_n\hat{A}_n+\hat{\xi}_n,
\end{equation} where $\hat{b}_n=\int dt \;\hat{b}(L_\mathrm{w},t)\psi^*_n(t)$ is the SFG output idler HG TM annihilation operator and $\hat{A}_n$ is a HG TM annihilation operator. The Schmidt mode noise annihilation operator is $\hat{\xi}_n=\int dt\;\hat{\xi}(t)\;\psi^*_n(t)$, which satisfies $[\hat{\xi}_n,\hat{\xi}^\dagger_n]=1-\lambda^2_n$ to ensure that $\hat{b}_n$ satisfies the bosonic commutation relations. As we have an expression of the SFG output idler channel annihilation operator in terms of the input signal HG TM annihilation operators we can proceed with calculation of the detection statistics of our QPMS LIDAR system. The HG TM annihilation operators will correspond to the quantum states derived in Appendix~\ref{sect:quantum_state}. For the numerical analysis of our system we set the decomposition coefficients (according to HG TM order) as follows: $\lambda_3=0.7$, $\lambda_4=0.07$, $\lambda_2=0.07$, $\lambda_5=0.014$, $\lambda_1=0.014$, $\lambda_6=0.014$ and $\lambda_0=0.014$. The decomposition coefficient for all other HG TMs is set to zero.
\label{sect:SFG}

\section{SFG Green function decomposition}
\label{sect:SFG_decomp}
As stated earlier, in general the Schmidt modes of the Green function will not be the HG TMs. Therefore, a further modal decomposition of the Green function Schmidt modes into the HG TMs of our quantum state is required. Following a procedure in Ref.~\cite{Lamata05} the Green function can be decomposed into a basis of HG TMs. From this HG decomposed form, a Schmidt decomposition is performed and the Green function for signal to idler in terms of its Schmidt modes is \begin{equation}
    G_{\mathrm{is}}(t,t^{'})=\sum^\infty_{n=0}\lambda_n \psi_n(t)\phi^*_n(t^{'}),\label{eq:gen_green}
\end{equation} where $\phi_n(t)$ is the input signal Schmidt mode, $\psi_n(t)$ is the output idler Schmidt mode and  $\lambda_n$ is the decomposition (Schmidt) coefficient. The Schmidt modes of Eq.\ref{eq:gen_green} will be some linear combination of the HG TMs and for numerical tractability the initial decomposition of the Green function into HG TMs is truncated. 

We can then decompose the signal to idler conversion relation (similar to Eq.~\ref{eq:input_output}) into a signal to idler conversion relation by Schmidt mode\begin{equation}
    \hat{b}_n=\lambda_n\hat{a}_n+\hat{\xi}_n,
\end{equation} where $\hat{b}_n=\int dt \;\hat{b}(L,t)\psi^*_n(t)$ is the idler Schmidt mode annihilation operator and $\hat{a}_n=\int dt^{'}\;\hat{a}(0,t^{'})\phi^*_n(t^{'})$ is the signal Schmidt mode annihilation operator. The Schmidt mode noise annihilation operator is $\hat{\xi}_n=\int dt\;\hat{\xi}(t)\;\psi^*_n(t)$, which satisfies $[\hat{\xi}_n,\hat{\xi}^\dagger_n]=1-\lambda^2_n$ to ensure that $\hat{b}_n$ satisfies the bosonic commutation relations. We then do a further modal decomposition of $\hat{a}_n$ into $\hat{A}_k$ (annihilation operators for HG TMs). Therefore, the input signal Schmidt mode is \begin{equation}
    \phi_n(t)=\sum_j V_{j,n}f_j(t),\label{eq:decomposed_phi2}
\end{equation}where $V_{j,n}=\int dt\;f^*_j(t)\phi_n(t)$ is the overlap integral and $f_j(t)$ is the TM amplitude of HG order $j$ in the time-domain. We can express the input signal Schmidt mode annihilation operator $\hat{a}_n$ in terms of the signal HG TMs $\hat{A}_j$ by substitution of Eq.~\ref{eq:decomposed_phi2} as\begin{align}
    \hat{a}_n&=\sum_j V^*_{j,n}\int dt^{'}\; f_j^*(t^{'})\hat{a}(0,t^{'}),\\
    &=\sum_j V^*_{j,n}\hat{A}_j.
\end{align}If $V_{j,n}=1$ for $j=n$ and $V_{j,n}=0$ for $j\neq n$, then our Schmidt modes are the HG TMs, and there is a one-to-one mapping from input signal HG TMs to output idler HG TMs.
\section{Photon-number statistics}
\label{sect:photon_number}
We detail the expectation values of the first two moments of the number operator $\hat{b}_n^\dagger\hat{b}_n$ for the signal-only state $\hat{\rho}_\mathrm{S}$ and the noise-only state $\hat{\rho}_\mathrm{B}$, from which we can calculate the SSMD. Moreover, we assume that the state corresponding to the noise operator $\hat{\xi}$ is in the vacuum and the Schmidt modes of the Green function are in the HG TM basis.

The expectation value of the first and second moment, respectively, of the number operator is \begin{equation}
    \bar{n}_{i:n}=\lambda^2_n\langle\hat{A}^\dagger_n\hat{A}_n\rangle
\end{equation} and \begin{equation}
    \langle\hat{b}_n^\dagger\hat{b}_n\hat{b}_n^\dagger\hat{b}_n\rangle=\lambda_n^4\langle\hat{A}^\dagger_n\hat{A}_n\hat{A}^\dagger_n\hat{A}_n\rangle+\lambda^2_n(1-\lambda^2_n)\langle\hat{A}^\dagger_n\hat{A}_n\rangle.
\end{equation}
Following on from this, for the signal-only state we have \begin{equation}
    \langle\hat{A}^\dagger_n\hat{A}_n\rangle=\eta\vert c^\mathrm{TM}_{n,(j=3)}\alpha\vert^2
\end{equation} and \begin{equation}
    \langle\hat{A}^\dagger_n\hat{A}_n\hat{A}^\dagger_n\hat{A}_n\rangle=\eta\vert c^\mathrm{TM}_{n,(j=3)}\alpha\vert^2(1+\eta\vert c^\mathrm{TM}_{n,(j=3)}\alpha\vert^2).
\end{equation} For the noise-only state we have \begin{equation}
    \langle\hat{A}^\dagger_n\hat{A}_n\rangle=\eta\vert c^\mathrm{TM}_{n,(j=0)}\alpha_c+\frac{\delta\alpha_c}{\sqrt{\eta}}\vert^2+\bar{n}_\mathrm{B,n}
\end{equation} and \begin{equation}
    \langle\hat{A}^\dagger_n\hat{A}_n\hat{A}^\dagger_n\hat{A}_n\rangle=\eta\vert c^\mathrm{TM}_{n,(j=0)}\alpha_c+\frac{\delta\alpha_c}{\sqrt{\eta}}\vert^2(4\bar{n}_\mathrm{B,n}+\eta\vert c^\mathrm{TM}_{n,(j=0)}\alpha_c+\frac{\delta\alpha_c}{\sqrt{\eta}}\vert^2)+2\bar{n}_\mathrm{B,n}^2+\langle\hat{A}^\dagger_n\hat{A}_n\rangle.
\end{equation}

\bibliography{specularbib} 
\bibliographystyle{spiebib} 

\end{document}